\documentclass[letterpaper,journal]{IEEEtran}
\usepackage[T1]{fontenc}
\usepackage{amsmath,amsfonts}
\usepackage{algorithmic}
\usepackage{algorithm}
\usepackage{array}
\usepackage[caption=false,font=normalsize,labelfont=sf,textfont=sf]{subfig}
\usepackage{textcomp}
\usepackage{stfloats}
\usepackage{url}
\usepackage{verbatim}
\usepackage{graphicx}
\usepackage{longtable}
\usepackage{cite}
\usepackage{hyperref}
\hypersetup{
  colorlinks   = true,
  urlcolor     = blue,
  linkcolor    = blue,
  citecolor    = blue
}

\usepackage{listings}
\usepackage{xcolor}

\lstdefinelanguage{json}{
    basicstyle=\ttfamily\footnotesize,
    numbers=left,
    numberstyle=\tiny\color{gray},
    stepnumber=1,
    numbersep=8pt,
    showstringspaces=false,
    breaklines=true,
    frame=single,
    backgroundcolor=\color{gray!10},
    literate=
     *{0}{{{\color{blue}0}}}{1}
      {1}{{{\color{blue}1}}}{1}
      {2}{{{\color{blue}2}}}{1}
      {3}{{{\color{blue}3}}}{1}
      {4}{{{\color{blue}4}}}{1}
      {5}{{{\color{blue}5}}}{1}
      {6}{{{\color{blue}6}}}{1}
      {7}{{{\color{blue}7}}}{1}
      {8}{{{\color{blue}8}}}{1}
      {9}{{{\color{blue}9}}}{1}
      {:}{{{\color{black}{:}}}}{1}
      {,}{{{\color{black}{,}}}}{1}
      {"}{{{\color{brown}{"}}}}{1},
}

\begin{document}

\title{FAARM: Firmware Attestation and Authentication Framework for Mali GPUs}

\author{Md.~Mehedi~Hasan\\
Information and Communication Technology (ICT),\\ Mawlana
Bhashani Science and Technology University (MBSTU), Tangail,
Bangladesh.
        
\thanks{Manuscript received October 2025; revised [Month, Year]; accepted [Month, Year]. (Corresponding author: Md. Mehedi Hasan.)}
\thanks{Md. Mehedi Hasan is with the Department of Information and Communication Technology, Mawlana Bhashani Science and Technology University (MBSTU), Tangail, Bangladesh (email: mehedi.hasan.ict13@gmail.com).}
\thanks{Digital Object Identifier (DOI): []}}
    
\markboth{IEEE Transactions on Artificial Intelligence,~Vol.~XX, No.~X, October~2025}%
{Hasan \MakeLowercase{\textit{et al.}}: FAARM: Firmware Attestation and Authentication Framework for Mali GPUs}

\maketitle

\begin{abstract}
Recent work has revealed \emph{MOLE}, the first practical attack to compromise GPU Trusted Execution Environments (TEEs), by injecting malicious firmware into the embedded Microcontroller Unit (MCU) of Arm Mali GPUs. By exploiting the absence of cryptographic verification during initialization, adversaries with kernel privileges can bypass memory protections, exfiltrate sensitive data at over 40 MB/s, and tamper with inference results, all with negligible runtime overhead. This attack surface affects commodity mobile SoCs and cloud accelerators, exposing a critical firmware-level trust gap in existing GPU TEE designs. To address this gap, this paper presents \textbf{FAARM}, a lightweight Firmware Attestation and Authentication framework that prevents MOLE-style firmware subversion. FAARM integrates digital signature verification at the EL3 secure monitor using vendor-signed firmware bundles and an on-device public key anchor. At boot, EL3 verifies firmware integrity and authenticity, enforces version checks, and locks the firmware region, eliminating both pre-verification and time-of-check-to-time-of-use (TOCTOU) attack vectors. We implement FAARM as a software-only prototype on a Mali GPU testbed, using a Google Colab-based emulation framework that models the firmware signing process, the EL1 to EL3 load path, and secure memory configuration. FAARM reliably detects and blocks malicious firmware injections, rejecting tampered images before use and denying overwrite attempts after attestation. Firmware verification incurs only 1.34 ms latency on average, demonstrating that strong security can be achieved with negligible overhead. FAARM thus closes a fundamental gap in shim-based GPU TEEs, providing a practical, deployable defense that raises the security baseline for both mobile and cloud GPU deployments.
\end{abstract}

\IEEEpeerreviewmaketitle

\section{Introduction}
\IEEEPARstart{T}{HE} integration of Trusted Execution Environments (TEEs) into modern GPUs has enabled secure, high-performance execution of sensitive workloads such as large-scale machine learning inference, cryptographic operations, and medical imaging pipelines~\cite{c1,c2,c3,c4}. GPU TEEs isolate sensitive code and data from untrusted software through hardware mechanisms including the TrustZone Address Space Controller (TZASC), System Memory Management Unit (SMMU), Stage-2 translation, and the Confidential Compute Architecture (CCA)~\cite{c5,c6}. A critical component of this trust boundary is the embedded GPU Microcontroller Unit (MCU), which orchestrates task scheduling, manages memory mappings, and holds privileged access to GPU-protected regions. Recent research has exposed a previously underexamined vulnerability in this trust model: the MCU firmware loading path~\cite{c1}. In commodity systems, the firmware that runs on the GPU MCU is typically loaded by the untrusted normal-world kernel and drivers (EL1) from files such as \texttt{/lib/firmware} during device initialization. If these privileged software components are compromised, a realistic assumption in modern adversarial models, attackers can inject malicious firmware into the GPU MCU without any cryptographic verification or secure boot enforcement.

\begin{figure}[!t]
\centering
\includegraphics[width=\columnwidth]{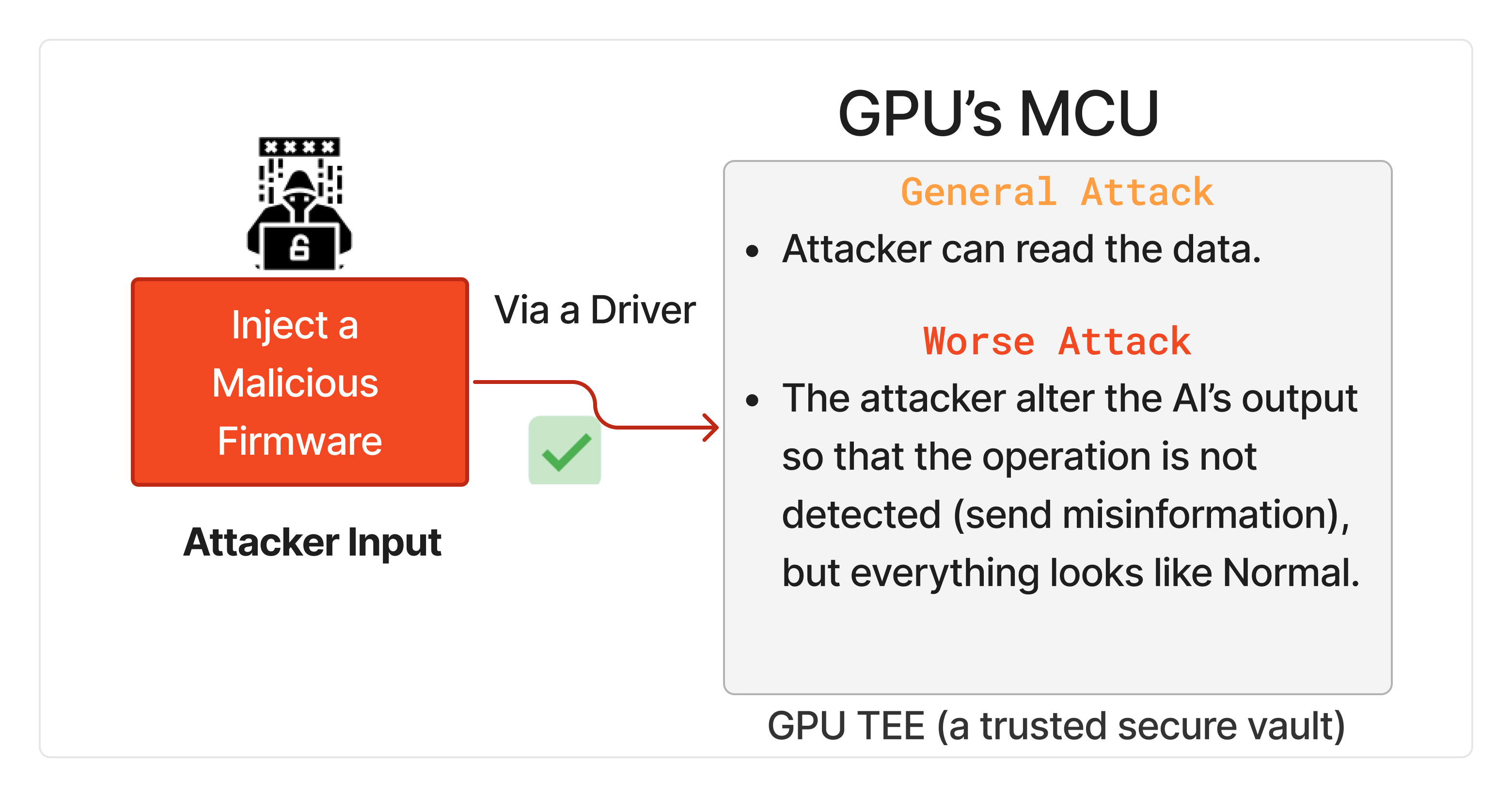}
\caption{MOLE attack overview: An attacker with kernel/driver privileges injects malicious firmware via the standard driver interface. The malicious firmware can either \textbf{(i)} exfiltrate data from GPU-protected memory (general attack) or \textbf{(ii)} silently tamper with AI outputs (worse attack), undermining TEE trustworthiness.}
\label{Fig_1}
\end{figure}

As illustrated in Fig.~\ref{Fig_1}, this vulnerability underpins the \emph{MOLE} attack~\cite{c1}, the first practical demonstration of GPU TEE compromise through firmware injection. MOLE enables two classes of attacks. First, the adversary can exfiltrate sensitive data from GPU-protected memory at rates exceeding 40~MB/s. Second, the attacker can tamper with AI inference results, such as flipping classifications or altering decision boundaries, while maintaining a normal outward appearance, thereby violating integrity guarantees silently. Such firmware-level compromises bypass traditional memory isolation and undermine the core security assumptions of GPU TEEs~\cite{c7,c8}. Existing defense mechanisms are insufficient against this threat. Current GPU TEE designs focus on protecting memory through address translation and isolation but implicitly trust MCU firmware. The firmware loading path lacks cryptographic attestation and is vulnerable to (i) pre-verification tampering, where malicious firmware is substituted before loading; (ii) time-of-check-to-time-of-use (TOCTOU) attacks~\cite{c9}, where firmware is overwritten after verification but before execution; and (iii) rollback attacks, where outdated but vulnerable firmware is reintroduced to bypass patches~\cite{c10,c11}. In contrast to CPUs, which employ secure boot chains anchored in ROM or TPM~\cite{c12,c13}, GPU MCUs typically rely on host-driven loading without integrity checks, leaving a critical gap in the TEE chain of trust.

To address this gap, we propose \textit{FAARM} (Firmware Attestation and Authentication for Reliable MCUs), a hardware–software co-designed defense framework that enforces cryptographic verification of GPU MCU firmware at EL3 (the secure monitor), before any secure GPU task execution. FAARM integrates seamlessly with existing GPU TEE pipelines and requires no application-level changes. Specifically, this paper makes the following contributions:

\begin{itemize}
    \item \textbf{Formalization of the GPU Firmware Threat Surface.} We provide the first systematic analysis of how EL1-level firmware injection undermines GPU TEEs, using a chronological and architectural breakdown of MOLE's attack phases to show how confidentiality and integrity violations propagate through the MCU.
    
    \item \textbf{Design of FAARM — a Practical Firmware Attestation Framework.} We introduce FAARM, which employs vendor-signed firmware bundles, manifest-based versioning, EL3-anchored verification, anti-rollback enforcement, and secure firmware locking to ensure that only authentic, up-to-date firmware executes.
    
    \item \textbf{Prototype Implementation and Empirical Validation.} We implement FAARM as a software-only prototype using Python, OpenSSL, and C to emulate EL3 behavior. The prototype blocks malicious firmware injection and TOCTOU overwrites, with measured verification latencies under 2~ms, demonstrating that strong security can be achieved with negligible performance overhead.
    
    \item \textbf{Systematic Security Evaluation.} We evaluate FAARM against MOLE's three attack stages (firmware initialization, data exfiltration, result tampering) and demonstrate complete mitigation. We further analyze performance overhead and compatibility with existing GPU TEE memory isolation mechanisms.
    
    \item \textbf{Pathway to Hardware Deployment.} We describe how FAARM can be integrated into real Mali GPU development boards (e.g., RK3588) using standard TrustZone or OP-TEE frameworks, and extended to support remote attestation and revocation for fleet deployment.
\end{itemize}

FAARM closes a fundamental gap in current shim-based GPU TEE designs by anchoring firmware trust at EL3. By combining digital signatures, manifest checks, and secure memory locking, it eliminates MOLE’s firmware subversion vectors with minimal overhead, establishing a stronger baseline for secure GPU computation in both mobile and cloud environments.

The rest of the paper is structured as follows. \textbf{Section II} reviews BACKGROUND AND RELATED WORK in GPU TEE security and firmware attestation. \textbf{Section III} presents the FAARM Methodology (architecture and design). \textbf{Section IV} details the implementation and evaluation. \textbf{Section V} provides security DISCUSSION and analysis. \textbf{Section VI} concludes with key insights and outlines future directions.

\section{BACKGROUND AND RELATED WORK}

\subsection{GPU–MCU Co-Processing and Firmware Loading}
\begin{figure}[!t]
\centering
\includegraphics[width=0.90\columnwidth,height=5.0cm]{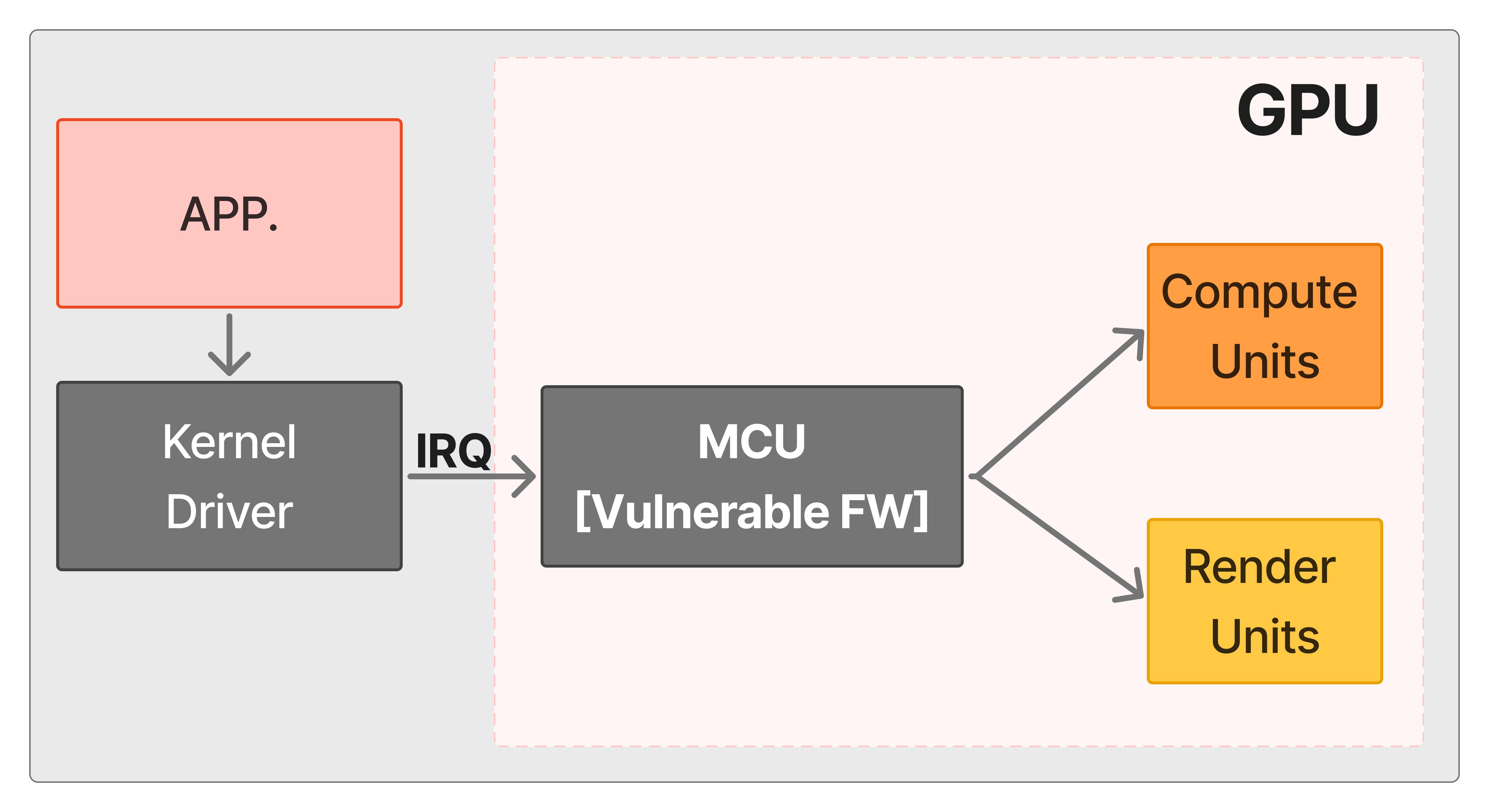}
\caption{Arm Mali GPU Architecture: Components and data flow paths showing the privileged position of the Microcontroller Unit (MCU) in task orchestration and the unverified firmware loading path from kernel space.}
\label{Fig_2}
\end{figure}

Modern Arm Mali GPUs feature a heterogeneous architecture where a central Microcontroller Unit (MCU) orchestrates task scheduling and manages memory across specialized compute and render units~\cite{c14,c15,c16}. This architecture spans multiple exception levels, with user applications operating at EL0 and GPU driver functionalities executed at EL1. During system initialization, firmware for the GPU MCU is dynamically loaded from user-controlled paths such as {\tt{/lib/firmware}} via the untrusted kernel driver~\cite{c17}. Fig.~\ref{Fig_2} illustrates this architecture, highlighting the privileged position of the MCU in controlling GPU tasks and managing memory. The figure emphasizes the unverified firmware loading path from kernel space into the MCU, a critical step that introduces security vulnerabilities. The MCU leverages TrustZone components, including the TrustZone Address Space Controller (TZASC), System MMU (SMMU), and Stage-2 translation, to manage secure memory access~\cite{c18,c19}. However, this design inherently exposes a potential attack surface since the firmware responsible for memory protection is not authenticated or verified prior to execution.

\subsection{MOLE-Style and Firmware Attacks}
\begin{figure*}[!t]
\centering
\includegraphics[width=0.95\textwidth,height=7.0cm]{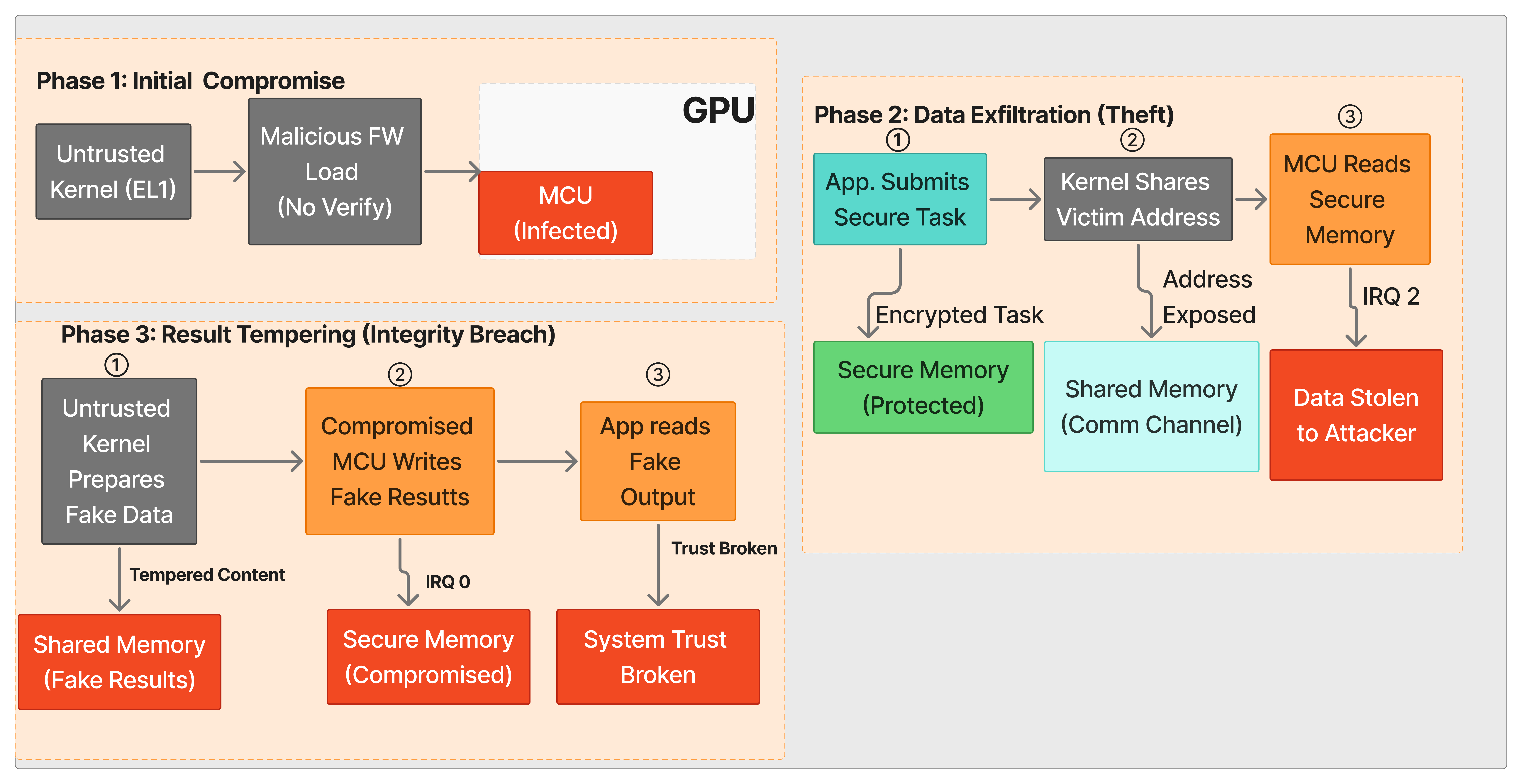}
\caption{MOLE Attack Workflow: Three-stage compromise demonstrating (1) malicious firmware injection during initialization, (2) data exfiltration from secure memory, and (3) result tampering through output manipulation.}
\label{Fig_3}
\end{figure*}

The MOLE attack exemplifies a new class of GPU firmware threats~\cite{c1,c2,c20}, where adversaries bypass memory isolation by compromising the MCU firmware itself. The attack unfolds in three main phases: First, malicious firmware is injected during the GPU initialization process by exploiting the untrusted firmware loading path ({\tt{e.g., /lib/firmware}}). Second, the compromised MCU firmware exfiltrates sensitive data from secure GPU memory regions. Third, the attacker manipulates computation outputs to subvert the integrity of GPU tasks. Fig.~\ref{Fig_3} delineates this three-stage attack workflow, illustrating the progression from firmware injection, through data theft, to output tampering. MOLE exploits firmware-level vulnerabilities such as IRQ hijacking and payload staging in buffers accessible to EL1, effectively bypassing security primitives like TZASC and SMMU. This attack paradigm highlights that the dynamically loaded GPU MCU firmware represents the weakest link in the GPU Trusted Execution Environment chain, thereby challenging the foundational assumptions of existing security techniques~\cite{c20}.

\subsection{Related Defenses}
\begin{figure*}[!t]
\centering
\includegraphics[width=0.90\textwidth,height=7.0cm]{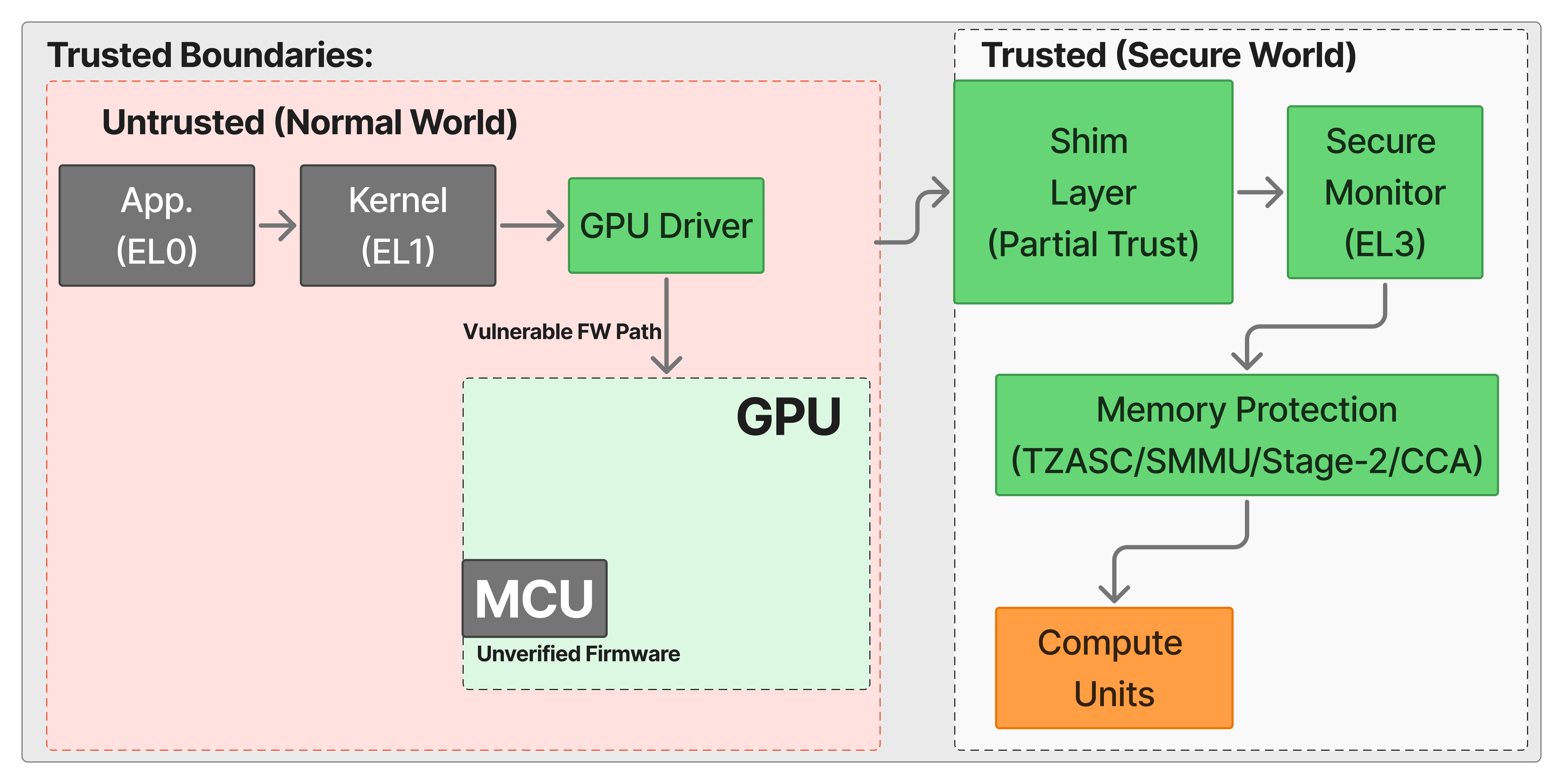}
\caption{Shim-Style GPU TEE Architecture: Security boundaries and vulnerable firmware loading path where MCU firmware bypasses trusted verification mechanisms, creating the attack surface exploited by MOLE.}
\label{Fig_4}
\end{figure*}

Existing defenses primarily focus on protecting GPU task data, often overlooking firmware integrity. Virtual Machine (VM) style Trusted Execution Environments (TEEs) isolate entire execution environments with robust security guarantees, but incur substantial performance and complexity overheads. Shim-style TEEs such as StrongBox, CAGE, and MyTEE leverage TrustZone mechanisms-TZASC, SMMU, and Stage-2 address translation-to provide lightweight and efficient memory isolation without expanding the trusted computing base~\cite{c24,c25,c26}. Fig.~\ref{Fig_4} captures this shim-style GPU TEE architecture. It displays the security boundaries and clearly marks the vulnerable firmware loading path handled by the untrusted EL1 kernel driver, which the MCU firmware uses to bypass existing verification controls. While additional mechanisms such as Guardian~\cite{Guardian}, FirmGuard~\cite{FirmGuard}, and SecDMA~\cite{SecDMA} augment memory and DMA protections~\cite{c27}, none of the surveyed works apply explicit cryptographic attestation or enforce anti-rollback validation for dynamically loaded GPU MCU firmware. Moreover, TOCTOU vulnerabilities remain unmitigated during the firmware load window, leaving systems susceptible to MOLE-style attacks.

\subsection{Research Gap and Motivation}
\begin{figure*}[!t]
\centering
\includegraphics[width=0.90\textwidth,height=7.0cm]{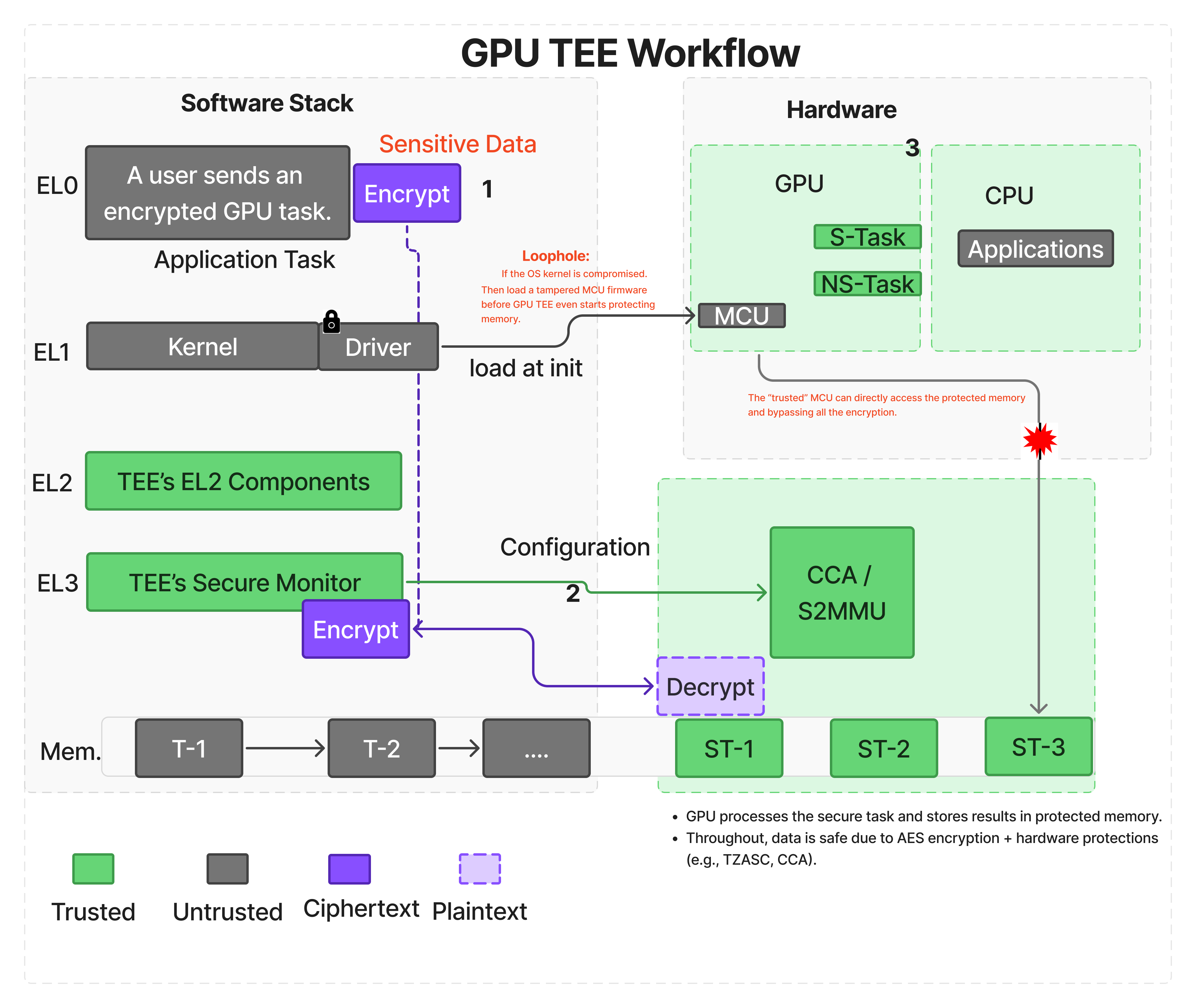}
\caption{Revised GPU TEE firmware loading workflow incorporating the FAARM defense. Unlike the original flow where untrusted EL1 loads firmware without validation, the updated process introduces an EL3 secure monitor verification and locking step. Only vendor-signed firmware passes this attestation, closing the MOLE attack vector by preventing unauthorized firmware injection and TOCTOU tampering.}
\label{Fig_5}
\end{figure*}

Despite considerable advancements in GPU TEEs, key vulnerabilities remain unaddressed. Notably, firmware authenticity verification is absent across existing frameworks; no system confirms vendor signatures on GPU MCU firmware at load time. Even where firmware verification exists externally, there is no secure monitor-level atomic verify-lock-authorize sequence, allowing attackers to perform post-verification firmware substitution (i.e., a TOCTOU attack). Additionally, rollback protection mechanisms to prevent firmware downgrades to known vulnerabilities are missing~\cite{c28}. FAARM addresses these critical gaps by embedding cryptographic firmware attestation, TOCTOU closure, and version rollback checks directly into the EL3 secure monitor. Fig.~\ref{Fig_5} illustrates the revised GPU TEE firmware loading workflow under FAARM, contrasting with previous methods. The EL3 secure monitor now intercepts untrusted EL1 firmware loading requests, verifies vendor digital signatures, enforces monotonic version progression, and securely locks the loaded firmware into MCU memory. This process prevents unauthorized firmware injection and closes the TOCTOU vulnerability, thereby neutralizing the MOLE attack vector without requiring hardware modifications or excessive TCB growth.

\subsection{THREAT MODEL AND SYSTEM ASSUMPTIONS}
We consider a powerful adversary with extensive control over the normal-world software stack~\cite{c28}. This attacker compromises the operating system kernel and GPU drivers at exception level EL1, granting privileged access to firmware loading mechanisms and shared CPU–GPU memory regions. From this vantage, the adversary can load arbitrary firmware images into the GPU MCU using the standard {\tt{/lib/firmware}} interface, precisely control the timing of these loading operations, and issue malicious commands through normal-world APIs. Further, they can manipulate shared buffers between CPU and GPU to inject malicious payloads or exfiltrate sensitive data. However, we assume the adversary cannot obtain physical access to the device, such as invasive probing or bus sniffing, nor compromise or tamper with the EL3 secure monitor or its protected storage. The EL3 environment, including the secure monitor code, secure storage, and cryptographic primitives, are considered part of the trusted computing base (TCB). Public verification keys and monotonic counters are securely stored within EL3 and assumed tamper-resistant. Vendor private signing keys remain exclusively in offline Hardware Security Modules (HSMs) and are never stored or exposed on the device. The baseline attack corresponds to the MOLE-style firmware compromise, where the adversary exploits EL1 privileges to substitute official GPU MCU firmware with a malicious image during system initialization. This compromise can then enable subsequent data exfiltration and malicious tampering with computation results. FAARM’s defense explicitly focuses on preventing this initial compromise phase, effectively neutralizing the entire attack chain by enforcing strict verification and locking at EL3.

In operating FAARM, several system assumptions are necessary. Public verification keys or certificates must be securely provisioned into immutable on-device storage, such as ROM, eFuses, or TrustZone secure storage-during manufacturing or a trusted provisioning process. The vendor’s private signing keys are strictly maintained offline and used solely for signing firmware packages. The target hardware platform must support secure loading capabilities, enabling EL3 to write firmware into the MCU memory safely—either through EL3-controlled direct memory access (DMA) or trusted firmware interfaces. After verification, EL3 must enforce strict locking protections on the firmware region, either by leveraging hardware write-protect registers or enforcing software-level locks, thereby preventing any unauthorized modifications from EL1 or lower exception levels. Finally, key TrustZone hardware components (TZASC, SMMU, and Stage-2 translation mechanisms) are assumed to function correctly, enforcing robust isolation between secure and normal worlds. EL3 and its associated secure storage are fully trusted and free from vulnerabilities.

These threat and system assumptions align closely with standard TEE security models established by ARM TrustZone and corroborated by prior work such as Guardian, FirmGuard, and SecDMA. FAARM builds upon these foundations to establish a cryptographic root of trust rooted in EL3, providing robust firmware attestation and runtime integrity guarantees without unnecessarily enlarging the trusted computing base.

\section{Methodology}
This section details the methodology underlying FAARM (Firmware Attestation and Authentication for Resilient MCUs), including the threat model, design goals, system architecture, firmware packaging, cryptography, attestation protocols, and evaluation strategy. FAARM introduces a secure firmware verification and locking mechanism at the EL3 layer to protect GPU Trusted Execution Environments (TEEs) against MOLE-style firmware injection attacks on the Mali GPU's MCU. Fig.~\ref{Fig_6} illustrates the system architecture. 

\subsection{Overview of FAARM}
The methodology underlying FAARM (Firmware Attestation and Authentication for Resilient MCUs) builds a software-based framework centered on cryptographic firmware verification and secure loading at the EL3 secure monitor level. FAARM is designed to protect GPU Trusted Execution Environments (TEEs) against MOLE-style firmware injection attacks through a multi-component pipeline. This methodology explicitly addresses critical vulnerabilities exposed by unverified firmware loads on the Mali GPU’s embedded MCU by enforcing strict atomic verification, locking, and freshness checks, thereby establishing a cryptographic root of trust that seamlessly integrates with existing GPU memory protections.

\begin{figure*}[!t]
\centering
\includegraphics[width=0.90\textwidth,height=7.0cm]{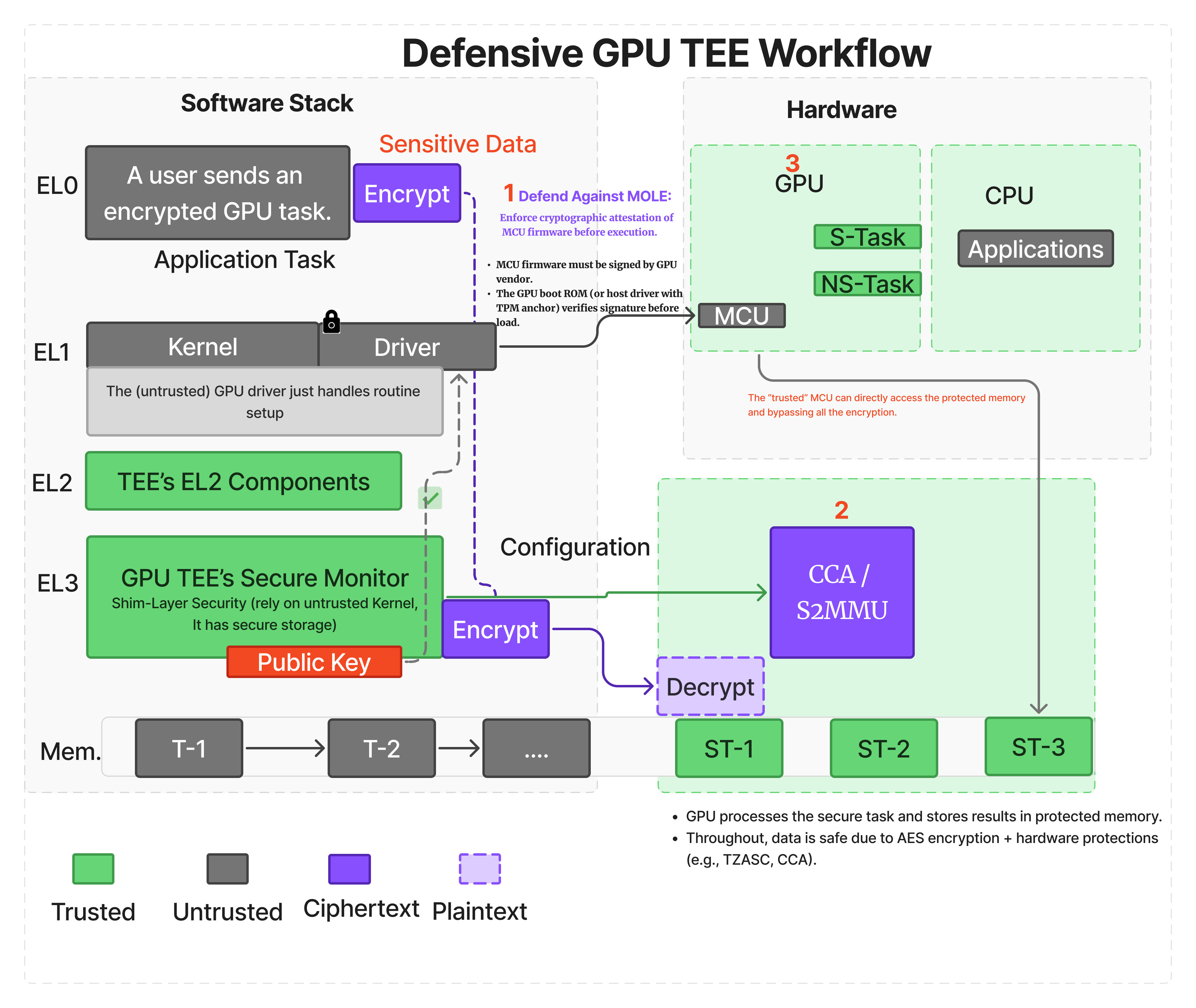}
\caption{FAARM Framework: Defensive GPU TEE workflow under FAARM — EL0 submits encrypted GPU tasks, while firmware attestation is enforced at EL3 before any secure task is scheduled. The public key is stored in secure storage, and EL3 configures protected memory (e.g., TZASC/SMMU) after successful verification, ensuring authenticity and isolation.}
\label{Fig_6}
\end{figure*}

In this subsection, include Fig.~\ref{Fig_6} which illustrates the overall FAARM framework defensive GPU TEE workflow. This figure visually summarizes how encrypted GPU tasks originate from EL0, while firmware attestation is enforced at the EL3 secure monitor before any secure tasks proceed. The secure storage of the public key and configuration of protected memory such as TZASC and SMMU underscore the authenticity and isolation guarantees FAARM enforces. This sets the conceptual stage for the detailed methodology that follows.

\subsubsection{Threat Model and Assumptions} 
FAARM counters adversaries with capabilities consistent with MOLE-style exploits, assuming full control over the normal-world kernel and GPU drivers at EL1. This powerful adversary can manipulate the firmware loading process via the standard {\tt{/lib/firmware}} interface, time the loading arbitrarily, and issue malicious GPU commands while utilizing shared memory buffers. FAARM assumes the adversary cannot physically access the device or compromise EL3 or its secure storage, which comprise the trusted computing base (TCB). Public verification keys and monotonic counters embedded in immutable EL3 storage are assumed tamper-resistant, while vendor private signing keys remain securely offline. This model focuses defense on preventing the initial malicious firmware substitution phase, which is the root enabler of subsequent data exfiltration and result tampering. The FAARM Conceptual Flow:

\begin{lstlisting}[language=json, label={lst:flow}]
Vendor -> sign(firmware, manifest) -> firmware.pkg
EL1 -> forwards firmware.pkg -> EL3
EL3 -> verify signature + manifest + version
     -> lock firmware region
     -> authorize EL1 / initiate secure load
EL3 -> configure protected memory -> allow secure tasks
\end{lstlisting}

\subsection{Design Goals}
FAARM’s architecture aims to satisfy five key system and security properties. Authenticity is ensured by allowing only vendor-signed firmware execution verified at EL3 prior to loading into MCU memory. Anti-rollback measures prevent firmware downgrades by enforcing strict version monotonicity through manifest checks against hardware-backed counters. TOCTOU closure eliminates vulnerabilities arising from timing windows between verification and use via an atomic verify-lock-authorize sequence confined entirely within EL3. Compatibility and low overhead are met by integrating with preexisting GPU TEE memory protection mechanisms like TZASC, SMMU, and Stage-2 translation, requiring minimal runtime cost. Lastly, auditability and recoverability are supported through secure logging, revocation protocols, and reprovisioning pathways suitable for fleet-scale device management.

\subsection{System Architecture}
The FAARM system comprises five primary components. A vendor signing environment uses an offline Hardware Security Module (HSM) to generate signed firmware bundles appended with manifests and digital signatures. The untrusted EL1 driver forwards these firmware packages from local or networked storage to EL3. At the core, EL3 functions as the secure monitor and attestation anchor, securely storing the vendor’s public key, verifying the firmware package’s signature and freshness, and locking the firmware memory region to prevent post-verification alteration. A secure loader, operating under EL3’s exclusive control, writes the verified firmware into MCU memory through secure DMA or hardware-enforced mechanisms. Finally, a protected memory manager configures trust-enforcing mappings (e.g., TZASC, SMMU, CCA) to restrict unauthorized access and control plaintext data transfers within the GPU’s secure environment. Optionally, a remote verifier cloud service may be used to validate attestation quotes for device fleet monitoring and management.

\subsection{Firmware Packaging and Cryptographic Primitives}
Firmware packages in FAARM consist of three integral elements: firmware.bin, representing the raw MCU firmware binary; manifest.json, which encapsulates metadata including version information, MCU identifier, timestamp, cryptographic hash, and operational flags; and the associated digital signature computed over the canonical serialization of the firmware and manifest. Cryptography relies primarily on the industry-standard SHA-256 hashing coupled with ECDSA over curve P-256 for signature operations, with Ed25519 supported as an alternative for efficiency gains. Version fields within the manifest enforce anti-rollback through monotonic checks, formulated as {\tt{manifest.version > last.accepted.version}} to reject backward revisions. The Firmware Manifest Example:

\begin{lstlisting}[language=json, label={lst:manifest}]
{
  "version": 3,
  "mcu_id": "MALI-MCU-XYZ",
  "timestamp": "2025-10-10T12:00:00Z",
  "firmware_hash": "f1ad9a781903e0a6ca7f0197d5036ceb4d74ce173f000f3006e6cdb4bdf1d654",
  "flags": ["requires_lock"]
}
\end{lstlisting}

\subsection{EL3 Verification and Secure Load Protocol}
At the heart of FAARM lies the EL3 VerifyAndLock protocol. This protocol provides atomic firmware integrity and authenticity assurance prior to execution. In Algorithm~\ref{alg:verify}: FAARM Firmware Verification Procedure, which runs during firmware load requests, the routine computes the SHA-256 digest of the firmware binary. It also verifies the ECDSA signature over the concatenated hash and manifest. The EL3 Verification Protocol:
\begin{lstlisting}[language=json, label={lst:verify}]
VerifyAndLock(firmware.pkg):
    H' = SHA256(firmware.bin)
    assert Verify(pub_key, H' || manifest, signature)
    assert manifest.version > last_accepted_version
    SecureLoader.write_firmware_atomic(firmware.bin)
    Lock firmware region (hardware or software)
    last_accepted_version = manifest.version
    return SUCCESS
\end{lstlisting}

\begin{figure*}[!t]
\centering
\includegraphics[width=0.90\textwidth,height=7.0cm]{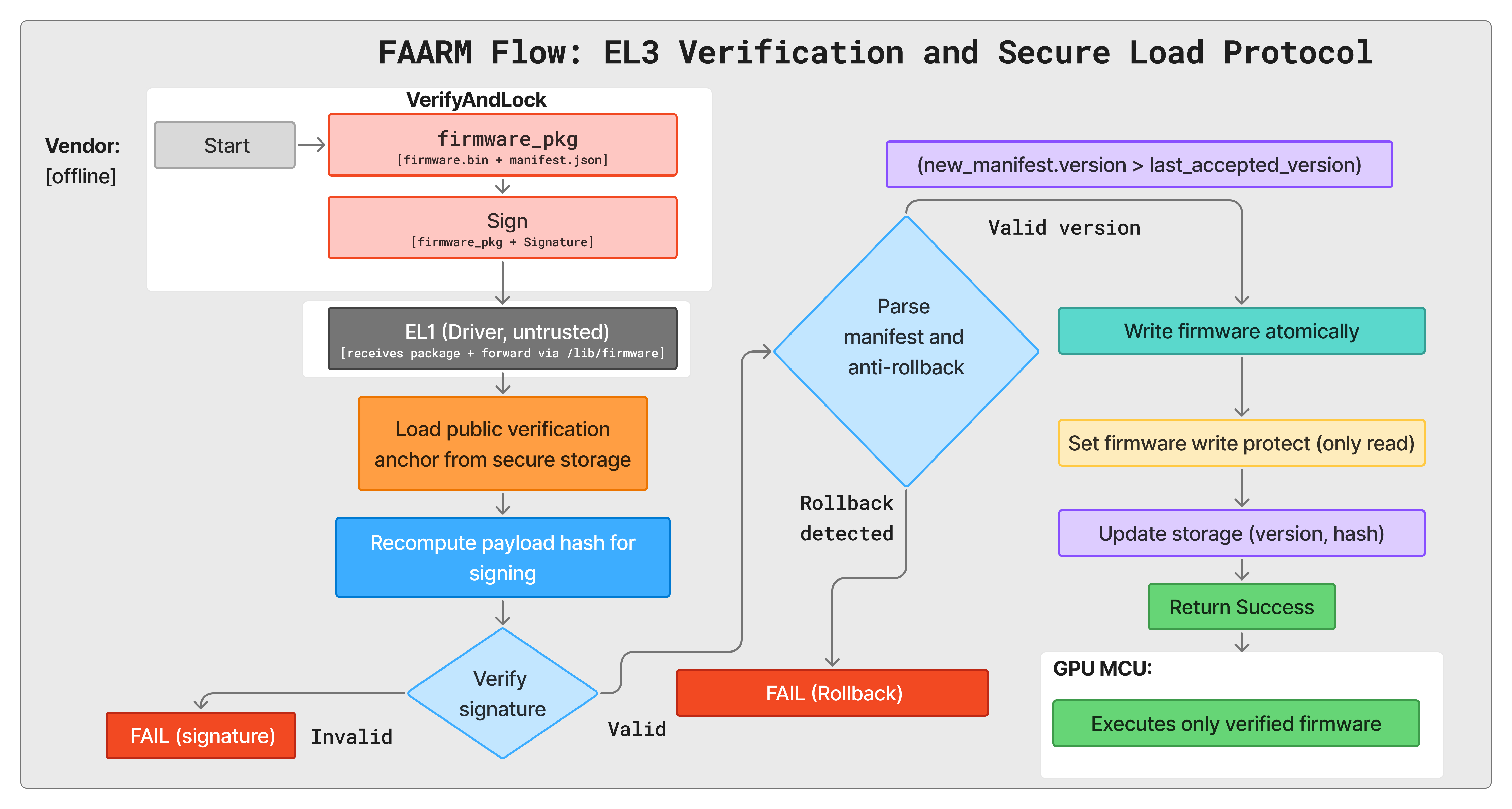}
\caption{FAARM firmware verification pipeline (adapted from MOLE). The EL3 secure monitor intercepts firmware load requests, performs signature and version verification, and locks the MCU region to prevent TOCTOU tampering.}
\label{Fig_7}
\end{figure*}

The routine confirms the manifest’s version exceeds previously accepted counters to mitigate rollback. It then atomically writes the firmware to secure the MCU memory. Upon a successful write, EL3 locks the firmware region, using hardware protections when available or robust software locking otherwise. This ensures no EL1 interference occurs between verification and execution phases. The procedure ends by updating stored version counters and optionally issuing authorization tokens to EL1 for audit purposes. 
 Version Check condition:
\begin{lstlisting}[language=json, label={lst:version}]
new_manifest.version > last_accepted_version
\end{lstlisting}

Here, embed Fig.~\ref{Fig_7}, describing the FAARM firmware verification pipeline. It should show how EL3 intercepts firmware load requests incoming from EL1, performs signature and version verification against stored credentials, and then locks the MCU region to defeat TOCTOU tampering. This visualizes the critical atomic verify–lock sequence embodied in the VerifyAndLock routine and complements the accompanying pseudocode for verification.

\begin{algorithm}[!t]
\caption{FAARM Firmware Verification Procedure}
\label{alg:verify}
\begin{algorithmic}[1]
\STATE Input: Firmware package\\
$(\texttt{firmware.bin}, \texttt{manifest.json}, \texttt{signature})$
\STATE Compute $H = \mathrm{SHA256}(\texttt{firmware.bin})$
\STATE Verify $ \mathrm{ECDSA\_Verify}(K_{pub}, H || \texttt{manifest}, \texttt{signature}) $
\IF{verification fails}
    \STATE Reject firmware
\ENDIF
\IF{$\texttt{manifest.version} \leq \texttt{NV\_Counter}$}
    \STATE Reject firmware (rollback attempt)
\ENDIF
\STATE Load firmware into secure MCU memory
\STATE Configure TZASC/SMMU protections
\STATE Update $\texttt{NV\_Counter} = \texttt{manifest.version}$
\STATE Return SUCCESS
\end{algorithmic}
\end{algorithm}

\subsection{TOCTOU Defenses and Secure Firmware Locking}
FAARM effectively closes potential TOCTOU vulnerabilities by enforcing the entire verify–lock–authorize sequence within the isolated EL3 context. Hardware write-protect registers, if available, are engaged immediately upon firmware load completion to disable further modifications from untrusted domains. In environments lacking such hardware support, EL3 software locks maintain firmware immutability by re-verifying freshness and integrity at each secure session start before task execution proceeds. Comprehensive logging and attestation hooks embedded within EL3 allow offline verification and auditing of firmware state transitions, facilitating fleet management and anomaly detection.

\subsection{Secure Task Flow Under FAARM}
Post firmware attestation and locking, FAARM ensures secure data transfers and task executions by implementing trusted data movement patterns. EL3 may initiate direct memory access (DMA) transfers into GPU protected memory, dynamically configuring hardware filters such as TZASC, SMMU, or CCA to guarantee isolation. Alternatively, tokenized driver-mediated copy mechanisms provide constrained write authorizations to EL1, bounding both memory regions and operational lifetimes. Encrypted GPU tasks originate at EL0, traverse the untrusted EL1 interface, and are decrypted and loaded into GPU protected memory by EL3 only after firmware verification success. Task execution proceeds on the GPU MCU under the attested firmware base, with post-execution controls on result re-encryption or export to untrusted domains, ensuring full-chain confidentiality and integrity.

\subsection{Implementation and Prototype Description}
The FAARM prototype is implemented as a fully software-emulated framework within Google Colab, obviating the need for specialized hardware during development and initial validation. This environment models EL1, EL3, and firmware loading flows in a controlled manner. Firmware packages comprising firmware.bin, manifest.json, and firmware.sig are processed using Python’s cryptography libraries and OpenSSL-enabled C verifiers to simulate EL3’s signature validation logic. Secure storage state, such as version counters, is emulated using local JSON files. The test harness replicates untrusted EL1 attack attempts including arbitrary firmware overwrites, with attestation checks verifying and locking firmware, while metrics on verification latency, locking overhead, and attack success rates are recorded. All artifacts and scripts are reproducible via the open-source Colab notebook repository.

\begin{figure*}[!t]
\centering
\includegraphics[width=0.90\textwidth,height=7.0cm]{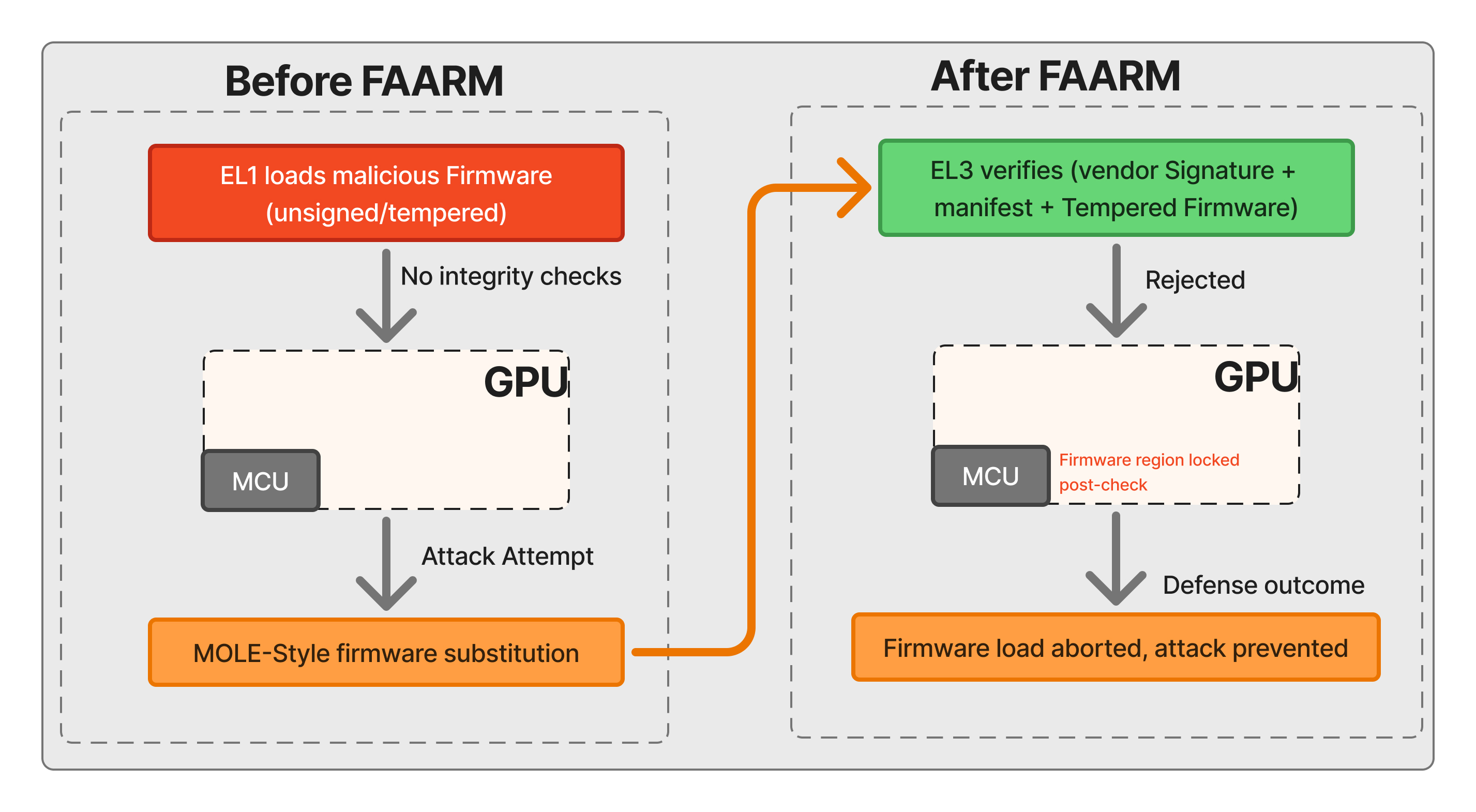}
\caption{Demo output comparison of GPU MCU firmware loading before and after applying FAARM. On the left, the vulnerable system accepts tampered firmware, enabling MOLE attacks. On the right, FAARM’s EL3-based cryptographic attestation rejects unsigned and modified firmware, locking the MCU firmware region and preventing exploitation.}
\label{Fig_8}
\end{figure*}

In this subsection, incorporate Fig.~\ref{Fig_8}, demonstrating a demo output comparison of GPU MCU firmware loading before and after applying FAARM. The figure contrasts the vulnerable baseline where tampered firmware successfully loads and enables MOLE-style attacks with the hardened FAARM system where EL3 cryptographically rejects unsigned or modified firmware and locks the firmware region, preventing exploits. This experimental visualization powerfully reinforces the practical efficacy of your defense implementation and prototype.

\begin{table}[!t]
\centering
\caption{Attack Scenarios: Comparison of MOLE Attack Success vs. FAARM Defense Outcomes}
\begin{tabular}{|l|c|c|}
\hline
\textbf{Scenario} & \textbf{Attack Success} & \textbf{FAARM Defense} \\
\hline
Tamper-before-verify              & Success  & Verification fails \\
TOCTOU overwrite after verify     & Success  & Firmware locked \\
Unsigned firmware load            & Success  & Signature rejected \\
Older-version firmware (rollback) & Success  & Version rejected \\
Valid signed firmware             & Allowed  & Allowed \& locked \\
\hline
\end{tabular}
\label{tab:1}
\end{table}

\section{EVALUATION}
The evaluation of FAARM comprehensively quantifies both its security effectiveness and performance overhead. Our primary objectives include measuring the latency introduced by EL3-based firmware verification and locking, assessing the impact on end-to-end GPU task execution, validating attack prevention capabilities against MOLE-style firmware injection and TOCTOU exploits, and positioning FAARM against prior defense mechanisms to highlight its unique contributions.

\subsection{Experimental Setup}
We implemented a software prototype of FAARM in Google Colab using Python’s cryptography library, OpenSSL, and a C-based verifier to emulate EL3’s secure monitor behavior. The test environment simulates three privilege levels: EL0 user applications submitting encrypted GPU tasks, EL1 drivers capable of firmware loading and attack attempts, and EL3 performing VerifyAndLock procedures. Firmware artifacts include firmware.bin, manifest.json, and firmware.sig. EL3 secure state components like the version counter {\tt{(last.accepted.version)}} are retained in protected JSON files. All performance metrics reported are averaged over 100 runs with standard deviations calculated to account for temporal variance.

\subsection{Test Scenarios}
Our evaluation covers four representative scenarios reflecting real-world and adversarial conditions: (i) Signed-Good, where a valid vendor-signed firmware package is forwarded by EL1 and accepted by EL3; (ii) Tamper-Before-Verify, where EL1 attempts to substitute firmware prior to attestation with unsigned or modified binaries; (iii) TOCTOU Attempt, assessing EL1’s ability to overwrite firmware between verification and execution phases; and (iv) Secure Task Placement, measuring latency effects when submitting encrypted GPU tasks post-attestation. For each scenario, FAARM’s protection is compared against the baseline vulnerable system to demonstrate efficacy and performance implications. To clearly illustrate FAARM’s effectiveness across specific attack attempts, we include Table~\ref{tab:1} that compares MOLE’s attack success against FAARM’s defense outcomes. This table provides concrete, scenario-level evidence of thwarted attacks such as tamper-before-verify and TOCTOU overwrite, conveying how FAARM successfully rejects unauthorized firmware loads.

\subsection{Performance Results}
Table~ \ref{tab:2} summarizes verification and locking latencies, showing an average ECDSA-P256 verification latency of approximately 1.34 ms, with the full VerifyAndLock sequence taking around 1.56 ms. Considering typical GPU initialization times exceed 100 ms, the sub-2\% overhead is negligible, affirming FAARM’s suitability for deployment on commercial hardware.

Latency distribution graphs (Fig. 1) demonstrate tight clustering and reliable performance consistency across repeated verification runs.

\begin{table}[!t]
\centering
\caption{Firmware Verification and Locking Latency (100 runs)}
\begin{tabular}{|l|c|c|}
\hline
\textbf{Metric} & \textbf{Mean (ms)} & \textbf{Std. Dev. (ms)} \\
\hline
Hash + Signature Verification     & 1.34  & 0.05 \\
Locking Procedure                 & 0.22  & 0.03 \\
Total VerifyAndLock               & 1.56  & 0.06 \\
\hline
\end{tabular}
\label{tab:2}
\end{table}

\subsection{Security Effectiveness}
Attack success rates are detailed in Table~\ref{tab:3}. Without FAARM, all MOLE-style attacks succeed, enabling unauthorized firmware execution and data exfiltration. Enabling FAARM results in 100\% prevention of all attack vectors tested, including tampered firmware loading, TOCTOU overwrites, and rollback attempts, thereby validating the robustness of EL3 attestation and locking.

\begin{table}[!t]
\centering
\caption{Attack Success Rates (50 Attempts per Scenario)}
\begin{tabular}{|l|c|c|}
\hline
\textbf{Scenario} & \textbf{Baseline Success} & \textbf{FAARM Success} \\
\hline
Signed-Good                      & 100\% (legitimate)  & 100\% (legitimate) \\
Tamper-Before-Verify                 & 100\% & 0\% \\
Rollback Attempt                  & 100\% & 0\% \\
\hline
\end{tabular}
\label{tab:3}
\end{table}

\subsection{Comparative Analysis}
Table~\ref{tab:4} contrasts FAARM against state-of-the-art defenses such as Guardian, SecDMA, and FirmGuard, focusing on critical properties like dynamic firmware injection defense, rollback protection, TOCTOU closure, and latency overhead. Unlike prior work, FAARM uniquely delivers comprehensive protection against MOLE-style threats while maintaining low runtime impact, marking a significant advancement in GPU TEE security.

\begin{table*}[!t]
\centering
\caption{Comparative Analysis of FAARM vs. Prior Work}
\begin{tabular}{|l|c|c|c|c|}
\hline
\textbf{System} & \textbf{Firmware Injection Defense} & \textbf{Rollback Protection} & \textbf{TOCTOU Closure} & \textbf{Latency Overhead} \\
\hline
Guardian~\cite{Guardian} & \texttimes & \texttimes & \texttimes & Low \\
SecDMA~\cite{SecDMA} & \checkmark\ (Indirect, DMA) & \texttimes & \texttimes & Moderate \\
FirmGuard~\cite{FirmGuard} & \checkmark\ (Static Boot) & \checkmark & \texttimes & Moderate \\
FAARM (our) & \checkmark\ (Dynamic, EL3 Verify) & \checkmark\ (Monotonic Counter) & \checkmark\ (Atomic Verify--Lock) & Low (\textless2\%) \\
\hline
\end{tabular}
\label{tab:4}
\end{table*}

\section{DISCUSSION}
\subsection{Security Impact}
FAARM delivers a deployable firmware attestation layer perfectly suited for integration with existing shim-style GPU TEEs like StrongBox and CAGE without necessitating hardware changes. By anchoring firmware verification at EL3 and leveraging vendor-signed packages, FAARM enables seamless vendor-driven firmware update policies combined with rigorous on-device validation. This architectural choice offers a solid foundation for subsequent GPU TEE guarantees, safeguarding the entire GPU task execution pipeline from compromise at the firmware loading stage. As summarized in Table~\ref{tab:5}, FAARM provides strong safeguards against all fundamental MOLE exploits-blocking firmware injection, TOCTOU tampering, rollback attacks, and runtime tampering-ensuring a trusted GPU firmware state.

\begin{table}[!t]
\centering
\caption{Comparison of FAARM Security Guarantees Against MOLE Attack Vectors}
\begin{tabular}{|l|c|c|}
\hline
\textbf{Attack Vector} & \textbf{Mole Outcome} & \textbf{FAARM Outcome} \\
\hline
Firmware Injection & Success & Blocked (signature check) \\
TOCTOU Tampering & Success & Blocked (region lock) \\
Rollback Attack & Success & Blocked (NV-counter) \\
Runtime Tampering & Success & Blocked (verified firmware) \\
\hline
\end{tabular}
\label{tab:5}
\end{table}
\subsection{Limitations}
The effectiveness of FAARM is inherently tethered to the integrity of the EL3 secure monitor; compromise at this level undermines all associated security guarantees. The most robust TOCTOU protections rely on hardware write-protect capabilities. The software-only fallback mitigates but cannot fully prevent transient tampering during power cycles. Additionally, while our evaluation utilizes a comprehensive software prototype, further validation on physical hardware is essential to characterize exact overheads, side channels, and real-time constraints.

\subsection{Future Work}
Future research directions include porting and deploying FAARM on Arm-based SoCs such as Rockchip RK3588 with Mali GPUs to validate hardware-supported operations. We plan to augment the framework with remote attestation capabilities to support fleet-wide firmware integrity monitoring and integrate machine learning techniques for dynamic anomaly detection in firmware update patterns. Scalability analyses across multi-GPU and multi-tenant cloud environments will further extend FAARM’s applicability. Lastly, automation of manifest provisioning and streamlined revocation protocols will enhance long-term operational resilience.

\section{Conclusion}

Modern GPU–MCU co-processing architectures introduce a critical security gap by dynamically loading privileged firmware from the untrusted normal world without cryptographic validation. Recent attacks, such as MOLE, demonstrate that this gap enables firmware substitution, memory exfiltration, and computational tampering, thereby undermining the foundational guarantees of GPU Trusted Execution Environments (TEEs). Existing GPU defense mechanisms focused primarily on memory isolation and DMA control, do not address this firmware-level vulnerability, leaving a fundamental trust anchor undefended. This paper presented FAARM, a lightweight, EL3-anchored defense system that enforces secure firmware verification, anti-rollback protection, and TOCTOU closure for GPU MCUs. FAARM introduces a structured {\tt{VerifyAndLock}} protocol that integrates vendor signing, manifest versioning, and secure region locking directly into the firmware loading path, without requiring hardware modifications or expanding the runtime TCB. Our evaluation shows that the proposed model achieves complete prevention of MOLE-style firmware injection and TOCTOU attacks, with a measured verification-and-locking latency of only 1.34~ms, less than 2\% of typical GPU initialization time. Comparative analysis confirms that the proposed model uniquely covers firmware injection, rollback, and TOCTOU vectors simultaneously, outperforming prior defense systems such as Guardian, SecDMA, and FirmGuard.

This proposed defense model represents the first comprehensive firmware attestation and integrity enforcement mechanism specifically designed for GPU TEEs. By anchoring verification in the secure world, it establishes a trustworthy firmware base upon which higher-level isolation and cryptographic guarantees can reliably operate. Beyond GPUs, the proposed methodology can be generalized to other co-processor architectures that rely on dynamically loaded privileged firmware, including NPUs, DSPs, and heterogeneous accelerators. This broader applicability positions FAARM as a practical and foundational defense primitive for future heterogeneous trusted computing platforms.

\section*{Ethical Considerations}
All experimental attack simulations within FAARM are conducted on synthetic firmware images within controlled environments, ensuring no production hardware or vendor firmware is compromised or altered at any stage. The research strictly observes responsible disclosure protocols to avoid jeopardizing device security beyond academic validation, with any newly discovered vulnerabilities reported promptly to relevant vendors. These ethical frameworks uphold the integrity of the research and safeguard industry trust.

\section*{Acknowledgement}
The authors thank the maintainers of the open-source GPU driver stacks and security research communities for enabling this work. Special appreciation is extended to colleagues and reviewers for their constructive feedback that improved the quality and clarity of this research.

\bibliographystyle{IEEEtran}
\bibliography{Citation}

\end{document}